%% file: 9407368.tex
\input mtexsis
\input epsf

\vsize 8.5truein
\preprint
\pagenumbers
\showsectIDfalse
\input rior.tex%
\input rio20.tex
\singlespaced 
\input rio21.tex
\nosechead{\tbf References}%
\singlespaced
\ListReferences%
\vfil\eject%
\bye

%% file: rior.tex
\referencelist\singlespaced
\reference{1} 
K. Rajagopal and F. Wilczek, Nucl. Phys. {\bf B379}, 395 (1993); 
{\bf B404}, 577 (1993)
\endreference
\reference{2}
J.D. Bjorken, K.L. Kowalski and C.C. Taylor, 
SLAC preprint SLAC-PUB-6109 (April 1993);
K.L. Kowalski and C.C. Taylor, Case Western Reserve Univ. preprint 
CWRUTH-92-6 (June, 1992), hep-ph/9211282;
A.A. Anselm and M.G. Ryskin,
Phys. Lett. {\bf B266}, 482 (1989);
J.-P. Blaizot and A. Krzywicki,
Phys. Rev. {\bf D46}, 246 (1992)
\endreference
\reference{3}
C. M. G. Lattes, Y. Fujimoto, and S. Hasegawa, 
Phys. Rep. {\bf 65}, 151 (1980);
G. Arnison {\it et al.}, Phys. Lett. {\bf 122B}, 189 (1983);
G. J. Alner {\it et al.}, Phys. Lett. {\bf 180B}, 415 (1986), 
Phys. Rep. {\bf 154}, 247 (1987);
J. R. Ren {\it et al.}, Phys. Rev. {\bf D 38}, 1417 (1988);
Chacaltaya Collaboration, in proceedings of the {\it 22nd International
Cosmic Ray Conference} (Dublin, Ireland) {\bf 4}, 89 (1991);
L. T. Baradzei {\it et al.}, Nucl. Phys. {\bf B370}, 365 (1992)
\endreference
\reference{4}
S. Gavin, A. Gocksch and R.D. Pisarski, Phys. Rev. Lett. {\bf 72}, 2143 (1994)
\endreference
\reference{5}
S. Gavin and B. M\"uller, Phys. Lett. {\bf B329}, 486 (1994)
\endreference
\reference{6}
S. Gavin, H. Satz, R.L. Thews, and R. Vogt, Z. Phys. \bf C61\rm\
(1994) 351; Nucl.\ Phys.\ {\bf A566}, (1994) 383c; J.-P. Blaizot, and
J.-Y.  Ollitrault, in `Quark-Gluon Plasma,' R. C. Hwa, ed. (World
Scientific, Singapore, 1990), pp. 631-663
\endreference
\reference{7}
F. R. Brown {\it et al.}, Phys. Rev. Lett. 65, 2491 (1990);
C. Bernard {\it et al.}, AZPH-TH-93-29, (1993);
S. Gavin, A. Gocksch and R.D. Pisarski, Phys. Rev. {\bf D49}, 307, (1994)
\endreference
\reference{8}
L. D. McLerran, Rev. Mod. Phys. 58 ( 1986) 1021-1064;
J. Cleymans, R.V. Gavai, and E. Suhonen, Phys. Rept. 130, 217 (1986)
\endreference
\reference{9}
J.F. Donoghue, E. Golowich, and B.R. Holstein, `Dynamics of the
Standard Model,' (Cambridge University Press, 1992)
\endreference
\reference{10}
D. Boyanovsky, H.J. de Vega, R. Holman, PITT-94-01 (1994), hep-ph/9401308
\endreference
\reference{11}
F. Cooper, Y. Kluger, E. Mottola, J.P. Paz, Los Alamos Preprint (1994),
hep-ph/9404357
\endreference
\reference{12}
G. Baym and G. Grinstein, Phys.\ Rev.\ \bf D15\rm\ 2897 (1977);
K. M. Benson, J. Bernstein, and S. Dodelson, Phys. Rev. {\rm D44}, 
2480, (1991)
\endreference
\reference{13}
D. Horn and R. Silver, Ann. Phys. (N.Y.) {\bf 66}, 509
(1971);
M. Gyulassy, S.K. Kauffmann, L.W. Wilson, Phys. Rev. {\bf C20} 
2267 (1979); 
R.D. Amado , F. Cannata, J.P. Dedonder, M.P. Locher, Bin Shao,
Phys. Rev. Lett. 72, 970 (1994)
\endreference
\reference{14}
See, {\it e.g.} W. Zajc, in NATO Workshop on Particle Production
in Highly Excited Matter, H. H. Gutbrod and J. Rafelski eds., (Plenum,
New York, 1993), p. 427
\endreference
\reference{15}
M. Gyulassy and S. S. Padula, Phys. Lett. {\bf B217}, 181 (1989)
\endreference
\reference{16}
S. Pratt, Phys. Lett. {\bf 301}, 159, (1993)
\endreference
\reference{17}
C. Greiner, C. Gong and B. M\"uller, Phys. Lett. {\bf B316}, 226 (1993)
\endreference

\endreferencelist

%% file: rio20.tex
\pubdate{July 1994}
\pubcode{BNL-60637}
\titlepage%
\title%
Dynamics of Chiral Symmetry Breaking in Nuclear Collisions
\endtitle%
\author
Sean Gavin
Department of Physics
Brookhaven National Laboratory
Upton, New York 11973--5000, USA
\endauthor
\abstract
Measurements of disoriented chiral condensates in heavy ion collisions
at RHIC and the LHC can yield fundamental information on the nature of
the QCD phase transition.  I review theoretical efforts to understand
the evolution of the condensate and present new results on
experimental signals in the single pion spectrum and in pion
interferometry.
\endabstract
\bigskip\bigskip\bigskip
\center
Invited talk at the Third International Workshop on 
Relativistic Aspects of Nuclear Physics, 
24-27 August 1993, Rio de Janeiro, Brazil.
\endcenter
\disclaimer{DE--AC02--76CH00016}
\endtitlepage

%% file: rio21.tex
\singlespaced
The order of the chiral restoration phase transition in high
temperature QCD is currently unknown.  If the transition is second
order, Rajagopal and Wilczek (RW) propose that the nonequilibrium
dynamics in ion--ion collisions at RHIC can generate transient domains
in which a macroscopic pion field develops.$^1$ These domains are
called disoriented chiral condensates (DCC).  Bjorken, Kowalski,
Taylor and others pointed out that DCCs can lead to fluctuations in
the charged and neutral pion spectra.$^2$  Hints of this behavior may have
been seen in the Centauro cosmic ray events.$^3$  In ion--ion collisions,
the ability of experimenters to identify DCCs amidst the background
produced by conventional particle production and scattering critically 
depends on the domains' size and energy content.

In this talk, I explore the RW mechanism for domain formation to
determine the size of domains in nuclear collisions, reporting on work
with Gocksch and Pisarski$^4$ and M\"uller $^5$.  I then briefly
speculate on phenomenological consequences of DCC formation.  [For a
discussion of $J/\psi$ suppression as presented at this meeting, see
Ref. [6].]

Equilibrium high temperature QCD manifests a chiral symmetry if the
light up and down quarks are taken to be massless.  However, a phase
transition occurs at a critical temperature $T_c\sim 140$~MeV in
which chiral symmetry is broken by the formation of a scalar $\langle
{\overline q} q\rangle$ condensate.  In the real world chiral symmetry
is explicitly broken at all temperatures by the few--MeV current quark
masses.  In that case the nature of the transition has yet to be
established.$^7$ [For reviews of lattice and finite temperature QCD,
see e.g. Refs. [8].]

Rajagopal and Wilczek pointed out that the chiral condensate can
become temporarily disoriented in the nonequilibrium conditions
encountered in heavy ion collisions.  Near $T_c$, the approximate
chiral symmetry implies that the scalar condensate is nearly
equivalent to a pion--like pseudoscalar isovector condensate $\sim
\langle {\overline q}\gamma_5{\vec\tau} q\rangle$, where $\vec\tau$
are the Pauli isospin matrices.  Consequently, domains containing a
macroscopic pion field can appear as the temperature drops below
$T_c$.  Such domains will eventually disappear as the system evolves
towards the true vacuum in which only the scalar condensate is
nonzero.  In the finite--sized heavy ion system, however, the evolving
domains can radiate pions preferentially according to their isospin
content.

The excitation of a pion field over a substantial fraction of the
collision volume can result in novel fluctuations in the number of
neutral and charged pions. 
Consider a single ideal domain in which the pion field is oscillating
along a fixed isospin direction $\vec\pi$.  Pion production from this
domain is proportional to the square of the field strength, so that
the fraction of neutral pions $f$ is
$$
f =
|\pi^0|^2/\sum_{0,+,-}|\pi^a|^2 \equiv \cos^2\;\theta.  
\EQN 1
$$
All orientations of $\vec\pi$ are equally likely, so that the
probability of a finding a fraction $f$ is 
$$
P(f)
\propto d(\cos\;\theta)/df \propto f^{-1/2}
\EQN 2$$
It is therefore most probable for a single domain to emit more charged
pions than neutral.$^2$

The detectability of domains in a nuclear collision depends on their
size.$^4$ If the interaction volume in a nuclear collision is dominated by
a single domain, then DCC formation can lead to a measurable isospin
asymmetry.  If a nuclear collision instead produces many uncorrelated
small domains, the spectrum of fluctuations would be gaussian, as
expected, {\it e.g.}, if the pions were produced independently.

To develop a more concrete picture of how the chiral condensate
evolves, we use the linear sigma model, in which the pion field is
coupled to a scalar $\sigma$ field that characterizes the scalar
condensate $^1$.  The fields interact through a potential
$$
V = \lambda({\vec\pi}^2+\sigma^2-v^2)^2/4 - H\sigma .
\EQN 3
$$
Chiral symmetry transformations correspond to $O(4)$ rotations of the
vector field $\Phi = (\sigma, \vec \pi)$.  For $H=0$ this
symmetry is spontaneously broken at zero temperature, with
$\langle\sigma\rangle = v$.  Current algebra implies that
$\langle\sigma\rangle = f_\pi = 93$~MeV.$^9$  The external field $H$
($\propto$ the light quark masses) breaks the $O(4)$ symmetry
explicity and gives the pions a small mass $m_\pi =
\sqrt{H/f_\pi} = 140$~MeV.  In comparison, the sigma mass
$m_\sigma\sim \sqrt{2\lambda v^2}$ is quite large, perhaps $\sim
600$~MeV.

Following Ref. [1], I will take this model with the parameters fixed
at $T=0$ as the basis for a Ginzburg--Landau description of the chiral
transition. For $H=0$, the transition would then be second order with
an $O(4)$ order parameter $\Phi$.  DCC domains would then be
well defined entities in the thermodynamic limit.  While there is
strictly no transition for $H\neq 0$, {\it transient} domains can form
as follows.

In the idealized `quench' scenario proposed by Rajagopal and Wilczek,
one assumes that the system initially has $\Phi = 0$, as appropriate
at high temperature, and follows the development of the system using
$T=0$ equations of motion derived from $V$.  The initial state of the
system is clearly unstable.  The system ``rolls down'' from the
unstable local maximum of $V({\Phi})$ towards the nearly stable values
with $|{\Phi}| = v$ (the symmetry breaking term $-H\sigma$ is
relatively small).  Field configurations with ${\vec\pi}\neq 0$
develop during the roll--down period.  The field will eventually settle
into stable oscillations about the unique vacuum $(f_\pi,\vec{0})$ for
$H\neq 0$, but oscillations continue until interactions eventually
damp the motion. 

To be more concrete, the linearized equations of motion for the Fourier
components of the pion field are:
$$
{d^2 \over dt^2}{{\vec\pi}_{\vec {k}}} \; = \; 
\left\{ \lambda v^2 - k^2\right\}
{{\vec\pi}_{\vec {k}}}.
\EQN 4
$$
Field configurations with $\langle\Phi\rangle=0$ and momentum $k < 
\sqrt{\lambda}v $ are unstable and grow exponentially; modes 
with higher momenta do not grow.  The $k=0$ mode grows the fastest,
with a time scale 
$$
\tau_{R} = \{\lambda v^2\}^{-1/2} \sim \sqrt{2}/m_\sigma.
\EQN 5
$$
The unstable modes grow for a time of order $\tau_R$ until $\Phi^2$
approaches $v^2$, after which the system oscillates about $\sigma =
f_\pi$.  

Significantly, the time during which growth can occur is quite short,
$\tau_R\sim 0.5$~fm for $m_\sigma \sim 600$~MeV.  Rajagopal and
Wilczek found that the power $\propto \langle{\pi^a}_{ \vec{k}}
{\pi^a}_{ -\vec{k}}\rangle$ in the low momentum pion modes 
grows when the exact classical equations of motion are integrated,
demonstrating that domains indeed form.  [Here $\langle \ldots
\rangle$ represents an average over the initial of the
fields, which are taken to have a thermal spectrum.]  What is not
clear, however, is whether the time during which the system is
unstable is sufficient for truly large domains to form.

Gocksch, Pisarski and I$^4$ studied the domain size by
numerically integrating the equations of motion in the quench scenario
and extracting the spatial correlation function:
$$
\langle\pi(\vec{x},t)\pi(0,t)\rangle \propto \int d^3k 
\langle{\pi_{-\vec{k}}}{\pi_{\vec{k}}}\rangle {\rm e}^{-i{\vec{k}\cdot
\vec{x}}},
\EQN 6
$$
where isospin labels are implicit.  This correlation function contains size
information since $\langle\pi(\vec{x},t)\pi(0,t)\rangle\approx 
\langle\pi(0,t)\rangle^2$ holds inside the domain.  The
numerical results in Ref. [4] can be understood by taking
$\pi_k(t)$ from the linearized equation (4) and assuming
that the fluctuations are initially thermal, so that 
$\langle\pi_{-\vec{k}}\pi_{\vec{k}}\rangle
\sim \{E(\exp{E/T_c}-1)\}^{-1}$ for $E = \{k^2 + m_\pi^2\}^{1/2}$.
A saddle--point integration yields $\langle\pi(\vec{x})\pi(0)\rangle
\propto {\rm e}^{-x^2/8\tau_Rt}$.  The domain size,
$R_D\sim 2\tau_R\propto m_\sigma^{-1}$, is set by the sigma mass. For
$m_\sigma = 600$~MeV, we find that domains are essentially pion sized.
Numerical simulations confirm this result.

The lesson drawn from the quench scenario is that large nucleus--sized
domains can occur only if the sigma mass is small.$^{4,10,11}$
However, it is quite possible that the effective $m_\sigma$ is reduced
in the high energy density heavy ion environment.  In fact, at a truly
second order phase transition we expect $m_\sigma$ --- the inverse
correlation length --- to vanish!  In that case, the parameters of the
effective potential for our Ginzburg-Landau model cannot be {\it
naively} taken from zero-temperature physics as we have done.

To explore the role of the medium in domain formation, M\"uller and I
studied the evolution of the condensate in the presence of a
nonequilibrium bath of quasiparticles.$^5$  We find that larger domains
are possible due to two effects, {\it i}) the reduction of $m_\sigma$
and {\it ii}) the ``annealing'' of the system, rather than quenching,
by the slowly expanding heavy ion system.  [Our use of the word
``annealing'' is meant to distinguish our scenario from RW's
``quench'' --- the similarity of these terms to those used in
condensed matter physics is in some respects misleading.]

I illustrate the role of mass reduction by considering a system near
equilibrium.  The system is described by an effective potential
$V_{\rm eff}$ that has the behavior shown in Fig.~1.$^{12}$
In the Hartree mean--field approximation, the change of $V_{\rm
eff}(\Phi)$ as a function of temperature is determined essentially by
the coefficient of $\Phi^2$ , {\it i.e.} the mass term.  Taking $H=0$,
one obtains a linearized equation analogous to (4) that
implies a growth rate $\tau_{R,\,{\rm eff}}^{-1} \propto m_\sigma^{\rm
eff}\propto (T_c^2 - T^2)^{1/2}$.  Growth is very slow near the
$T_c\approx\sqrt{2}v$, prolonging the time over which the system is
unstable.  Only as the potential approaches its free space shape does
the roll--down become rapid.
\midfigure{1}
\epsfxsize=3.2in
\centerline{\epsffile{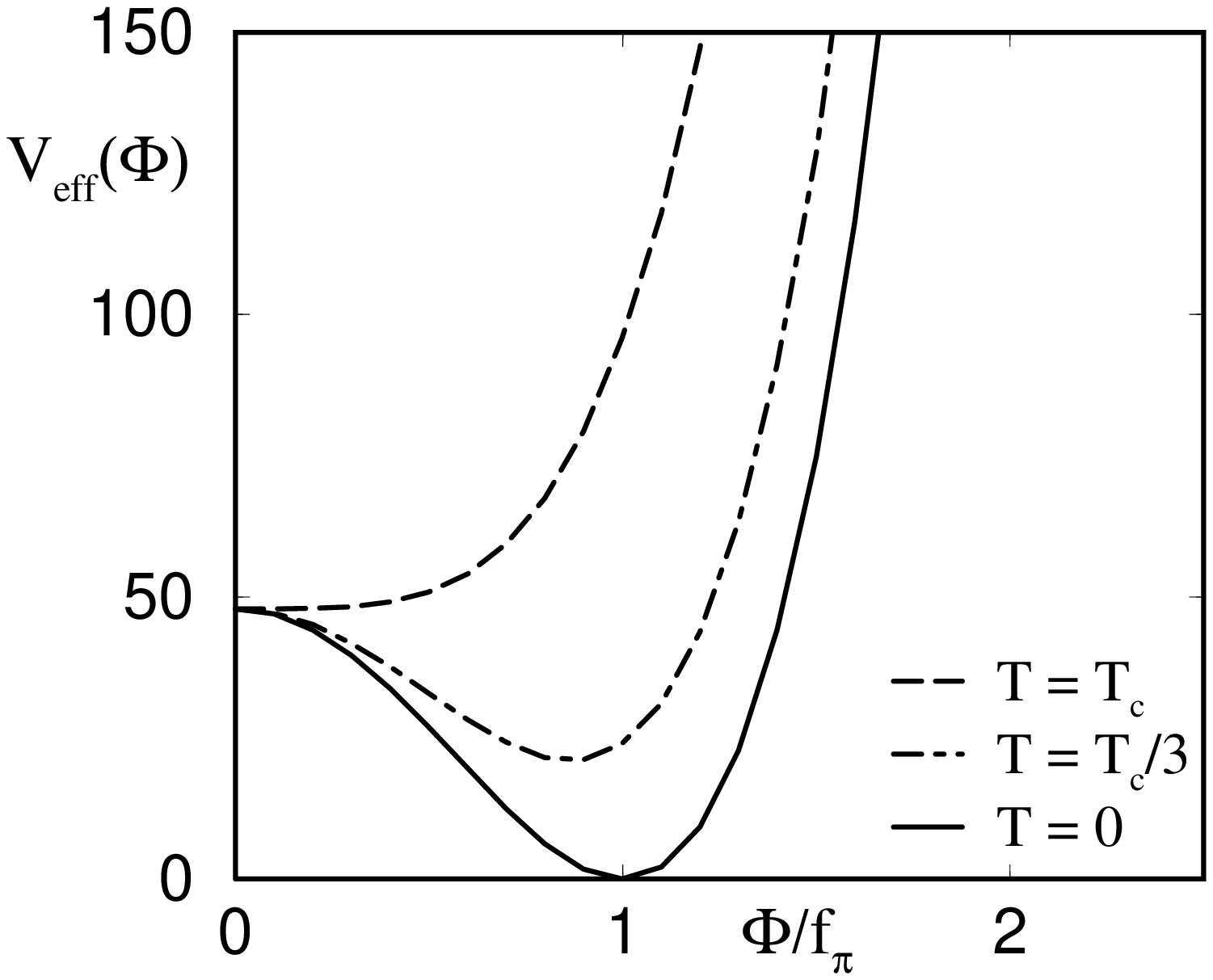}}
\Caption
The effective potential for various temperatures.
\endCaption
\endfigure

The annealing scenario describes the nonequilibrium evolution of the
system in a regime where the time scale $\tau_{\rm ev}$ for the
evolution of $V_{\rm eff}$ is much longer than $\tau_R$.  One expects
such a slow cooling scenario to further enhance the domain size --- in
metallurgy, the more rapidly one quenches, the smaller are the
resulting crystals.  The evolution of $V_{\rm eff}$ is determined by
the expansion of the surrounding quasiparticles, which constitute a
time--varying nonequilibrium heat bath.  The annealing regime is
relevant at RHIC energies because the temperature falls to $T_c\approx
140$~MeV only at very late times, perhaps $\tau_c\sim 10-20$~fm.  The
expansion time $\tau_{\rm ev}$ is then $\sim\tau_c \gg\tau_R\sim 0.5$~fm.

M\"uller and I explored the evolution in the annealing regime using a
nonequilibrium Hartree--like approximation. [See Ref. [5] for
details.]  We indeed find that annealing prolongs the time during
which the system is unstable compared to a quench.  Domain sizes up to
7 fm are possible if we assume that $m_\sigma^{\rm eff}(T_c) = 0$
({\it i.e.}, a second order transition).  More conservatively, we find
sizes of 3---4 fm for $m_\sigma^{\rm eff}(T_c) = 300$~MeV.  Such large
masses are possible, {\it e.g.}, in the large $N$ limit of the $O(N)$
model studied by Boyanovsky {\it et al.}$^{10}$ and Kluger {\it et
al.}$^{11}$  However, QCD may be far richer than the large $N$ limit implies.

Suppose that a nuclear collision produces a large domain, either in a
typical event due to the annealing dynamics or due to a rare
fluctuation.  How can we tell?  The number of pions is not large, even
ideally.  The amount of energy available for pion production per unit
volume is the potential difference $\Delta V \sim 60$~MeV~fm$^{-3}$
between $\Phi=0$ and $\Phi=(f_\pi,\vec{0})$.  If all of the energy in
a domain of size $R_D\sim 3-6$~fm goes into pion production, one
expects $N_{\rm dcc}\sim\Delta V R_D^3/m_\pi \sim 30 - 300$ pions.  In
comparison, conventional production mechanisms in a central Au+Au
collision at RHIC can produce 1000 pions {\it per unit rapidity}; this
constitutes a background to the DCC signal.

The source of pions from this domain is coherent and, therefore,
concentrated at low transverse momenta.  Horn and Silver, Gyulassy
{\it et al.}, and Amado {\it et al.} have developed a coherent state
formalism applicable to DCC production.$^{13}$  The amplitude for
emitting a single neutral pion from a space time point $(x_j, t_j)$
within a domain is real and satisfies $A_j\propto\pi^0(x_j)\propto {\rm
exp}\{-x_j^2/2R_D^2\}$.  The production rate from
a single domain is 
$$
E({{dN}/{d^3p}})_{\rm dcc}
\propto|\sum_j A_j {\rm e}^{- ip\cdot x_j + iE_{\vec p}t_j}|^2
\sim {\rm e}^{-p^2R_D^2}, 
\EQN 7
$$
neglecting an energy--dependent factor that is slowly varying for small
momenta.  The rate is largest for $p_T < R_D^{-1}$.

The DCC signal (6) appears atop an isospin symmetric background.
Assuming that all the DCC pions are neutral and taking the background
produced by the lund/fritiof event generator for central Au+Au, I
obtain the spectra in Fig.~2.  A signal that is significant compared
to statistical fluctuations in the background emerges at low $p_T$ for
a domain of size $R_D>3$.
\midfigure{2}
\epsfysize=3.2in
\centerline{\epsffile{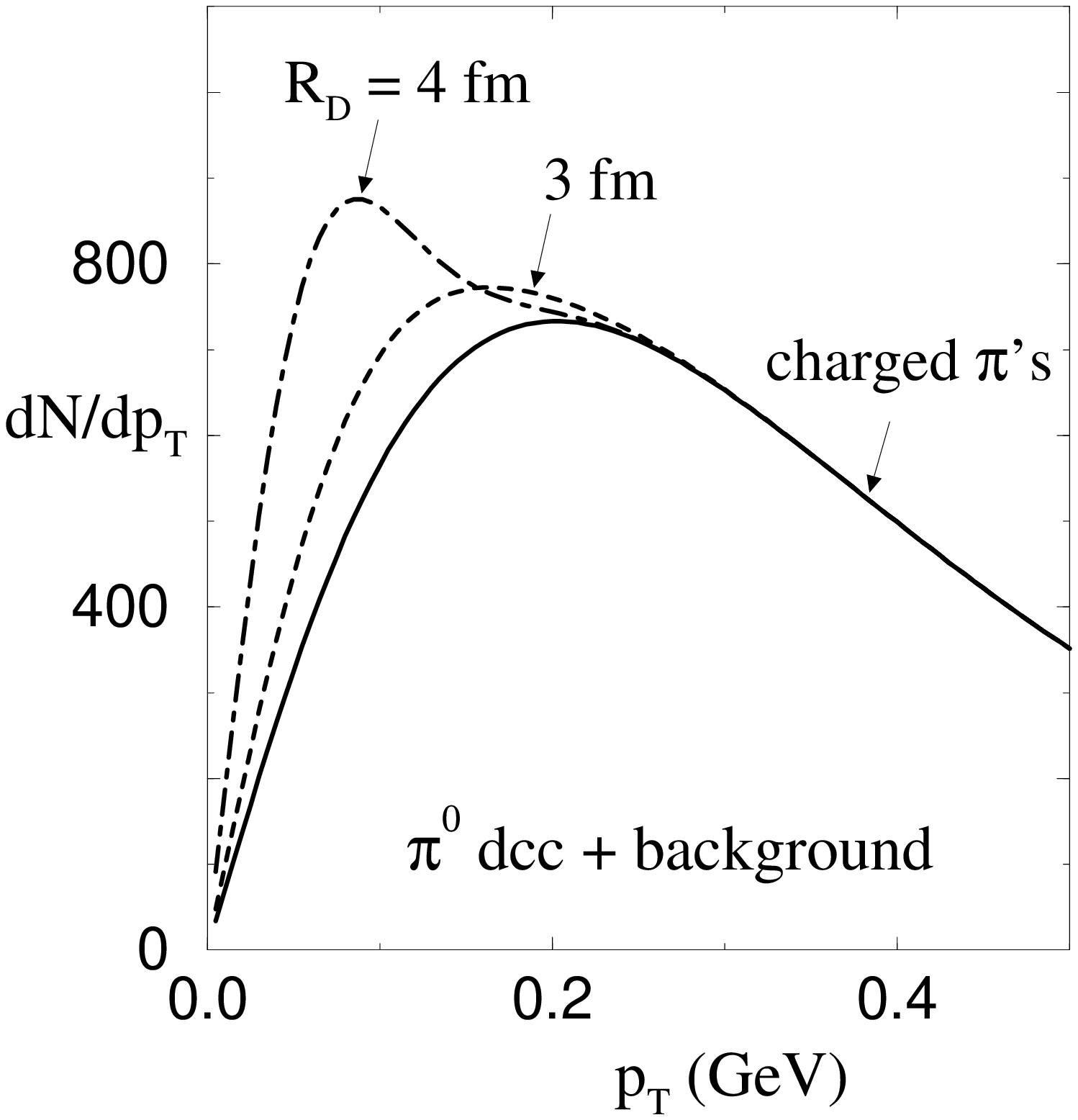}}
\Caption
Neutral pion production from a single large DCC in Au+Au at RHIC.
\endCaption
\endfigure

Experimentally, one must search for DCCs on an event--by--event basis,
looking for significant fluctuations in the neutral and
charged pions.  A domain producing a neutral fraction $f$ introduces a
contribution $fE(dN/d^3p)_{\rm dcc}$ to the $\pi^0$ spectrum, with a
corresponding contribution $\{(1-f)/2\} E(dN/d^3p)_{\rm dcc}$ to the
$\pi^+$ and $\pi^-$ spectra.  The distribution of $f$ in a sample of
large--domain events is given by (1).    

To check that the pion emission from a candidate large--domain event
is coherent, one can study the Bose correlations of pairs of identical
pions.  Pions produced by conventional scattering mechanisms are
largely incoherent, and exhibit an intensity interference that is
analogous to the Hanbury--Brown--Twiss effect for photons.$^{15}$
Measurements of $\pi\pi$ correlations are typically used to study
information on the size of the interaction volume in nuclear
collisions, much as the HBT effect is used to measure the size of
stars (although interpretation of experiments is highly
nontrivial$^{16}$).  In contrast, a fully coherent source of pions
exhibits interference at the amplitude level.  Consequently, the
single particle spectrum (7) from an isolated domain depends on the
size of the domain, while the pair distribution is simply the product
of single particle spectra.

We therefore expect the HBT effect to be suppressed in a nuclear
collision if a large domain forms.  However, the HBT effect does not
completely disappear, since intensity interference can occur when a
pion from the DCC has a similar momentum to a background pion.  To
estimate this effect, 
I follow Gyulassy {\it et al.} in Ref. [13] and write the two pion 
correlation function
$$\eqalignno{
C(p_1,p_2) = 1 &+ (1-D(p_1))(1-D(p_2)){\tilde\rho}(q)^2\cr
               &+ 2\{ D(p_1) D(p_2) (1-D(p_1)) (1-D(p_2))\}^{1/2}
{\tilde\rho}(q),&(8)\cr}
$$
where $q=p_1-p_1$ is the relative momentum of the pair and
$\tilde\rho$ is the Fourier transform of the space--time density of the
background pions.  Observe that (8) depends strongly on the fraction
of coherent pions $D = (dN/d^3p)_{\rm dcc}/(dN/d^3p)_{\rm tot}$.

The possible impact of a DCC on $\pi^-\pi^-$ correlations is
illustrated in Fig.~3 as a function of domain size.  I assume that the
domain produces only charged pions and that the transverse source size
for the fritiof background is 7~fm.  A combination of one-- and
two--particle measurements can allow one to disentangle the DCC signal
from the background.      
Of course, the interpretation of experiments
will be extremely tricky!
\midfigure{3}
\epsfxsize=2.8in
\vskip -1.2truein
\centerline{\epsffile{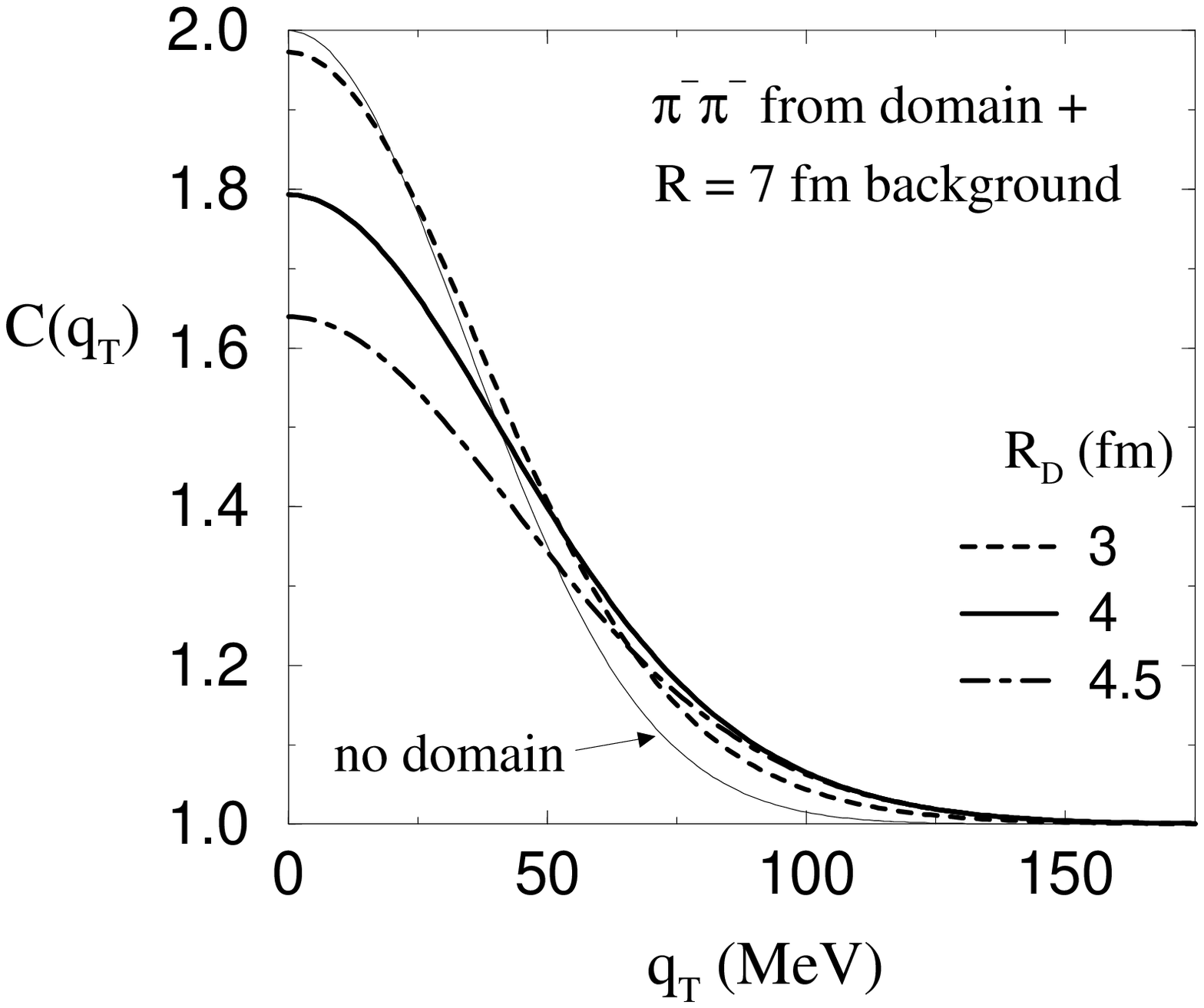}}
\Caption
$\pi^-\pi-$ correlations as a function of $q_T = p_{T1} - p_{T2}$ 
from a single large DCC plus a background of incoherent pions in 
Au+Au at RHIC. The total momentum of the pair is taken 
to be $K = (p_{T1} + p_{T2})/2 \equiv 0$.
\endCaption
\endfigure

The schematic results in Figs. 2 and 3 indicate that domains larger
than 3~fm can have measurable consequences.  If seen in experiments,
what would they tell us about QCD?  All of these signals --- the
isospin fluctuations, the enhancement of the pion spectrum at low
$p_T$, and the suppression of HBT correlations --- are characteristics
of any large coherent source.  In principle, Bose condensation in an
ideal pion gas can produce similar signals.$^{16}$ While it is
argued$^{17}$ that the conditions at RHIC are not favorable for Bose
condensation in the absence of a potential such as (3), systematic
experimentation will nevertheless be needed to prove that the
coherence comes from a disorientation of the chiral condensate.  If an
annealing scenario is valid, I expect large domains to occur in most
central collisions.  Lego plots$^2$ could then reveal a `clumpy' event
structure indicative of different domains.  The pion excess within each
domain would be independent of the background multiplicity.   On the
other hand, the number of condensed pions in ideal Bose condensation
would grow with the overall multiplicity and have a homogeneous
structure.

I am grateful to Andreas Gocksch, Rob Pisarski, and Berndt Mueller for 
their very enjoyable collaboration.